# Dual Plasmonic Modes from In-Situ Silver Nanoparticle Formation via Diffusion at Silver / Dielectric Interfaces

*Yufeng Ye[1,2], Joel Y.Y. Loh[1], Andrew Flood[1], Cong Y. Fang[1,2], Joshua Chang[1,2], Ruizhi Zhao[1,2], Peter Brodersen[3,4], Nazir P. Kherani[1,5]* \*

[1]Department of Electrical and Computing Engineering, University of Toronto, 10 King's College Road, Toronto, ON M5S 3G4, Canada

[2]Division of Engineering Science, University of Toronto, 40 Saint George Street, Toronto, ON M5S 2E4, Canada

[3]Department of Chemical Engineering and Applied Chemistry, University of Toronto, 200 College St, Toronto, ON M5S 3E5, Canada

[4]Ontario Centre for the Characterisation of Advanced Materials, University of Toronto, 200 College St, Toronto, ON M5S 3E5, Canada

[5]Department of Materials Science and Engineering, University of Toronto, 140 College Street, Toronto, ON M5S 3E4, Canada

*Correspondence: kherani@ecf.utoronto.ca; Tel.: +01.416.946.7372

**Abstract:** Metal-dielectric interfaces of various geometries are fundamental photonic material platforms for surface plasmons. Surface plasmon polaritons and localized surface plasmons are two surface plasmon modes that are excited on planar and curved metal-dielectric interfaces, respectively. Herein, we demonstrate in-situ formation of silver nanoparticles by diffusion at a planar interface between sputter-deposited silver and nitride dielectrics. In one step, we synthesize a nanostructure that exhibits both localized surface plasmon resonances at silver nanoparticles and surface plasmon resonances at the planar interface. We develop an accurate optical model describing both surface plasmon modes by applying effective medium theories to experimental ToF-SIMS and XPS depth profiles. Our findings provide fundamental material insights into intrinsic metal-dielectric interfacial defects, along with a new in-situ nanoparticle synthesis method that seamlessly integrates with conventional fabrication of planar interfaces. These unique results open the prospect of promising photonic-plasmonic applications availing the coupling of both surface plasmon modes.

**Introduction**

The metal-dielectric interface is a fundamental photonic material platform for exciting collective electron oscillations known as surface plasmons[1]. Depending on the geometry of the interface, a variety of resonant modes of the surface plasmon will emerge; two useful modes that have been extensively studied are the surface plasmon polariton (SPP), which propagates along planar metal-dielectric interfaces, and the localized surface plasmon (LSP), which is confined on subwavelength nanostructures such as metal nanoparticles in a dielectric medium[1–3]. The SPP finds important applications in metamaterials[4,5], sub-wavelength optics[6,7], biosensing[8], and super-lens[9,10], whereas the LSP offers higher field enhancement which is more apropos for surface-enhanced Raman spectroscopy[2], enhanced absorption[11], and non-linear applications[12,13].

Because of the strong geometrical dependence of surface plasmon modes, in-depth material morphology studies of metal-dielectric interfaces are crucial for modelling the optical performance of devices. For the planar interface, surface roughness and diffusion are two important defects identified in the literature. Many studies[14–17] have measured additional optical absorption from surface plasmon resonance (SPR) associated with the surface roughness of the metal film. In addition, controlled-environment diffusion studies[18,19] demonstrated that Ag, a standard plasmonic metal, can be highly diffusive when the interfaces undergo post-deposition treatments such as annealing[18] or UV photoillumination[19]. However, to the best of our knowledge, no study thus far has identified the morphology of the diffused silver. As such, the plasmonic effect of the diffusion phenomenon remains completely unexplored.

Herein, we investigate planar metal-dielectric interfaces between sputter-deposited Ag and two reactively sputter-deposited nitride dielectrics, namely, hydrogenated aluminum nitride (AlN) and silicon nitride ($SiN_x$). We find that Ag spontaneously diffuses into the nitride layer and forms nanoparticles during sputter deposition. Without any post-deposition treatment, the discovered phenomenon is intrinsic to the fabrication process and represents a facile in-situ method to synthesize Ag nanoparticles within the dielectric and the contiguous planar metal-dielectric interface using the conventional sputter-deposition method. We show that the embedded nanoparticles excite the characteristic localized surface plasmon resonance (LSPR), in addition to the well-known SPR due to surface roughness. These findings demonstrate a one-step procedure for fabricating a nanostructure that support both SPP and LSP modes, which are ideal for applications that combine[1] SPR and LSPR such as ultra sensitive immunoassays[20], DNA detection[21], and enhanced surface plasmon-coupled emission[22].

**Plasmonic absorption from Ag nanoparticles at AlN / Ag**

We begin by first focusing on AlN as the dielectric interfaced with Ag, a later section will generalize these results to SiN$_x$. Table 1 details the AlN / Ag / AlN multilayered samples fabricated with controlled variations in top AlN thickness and Ag layer thickness. The intense golden coloration of the films (Fig.1m) was immediately visible following sputter deposition of the samples and persisted for at least 14 months. UV-Vis absorption spectra of these samples (Fig. 1j,k) reveal the strong absorption peaks in the violet wavelengths (380-450 nm) which give rise to the observed complementary golden color[15].

**Table** 1. All fabricated thickness combinations of AlN / Ag / AlN thin film stacks on glass and silicon substrates. Samples S1, S2, S3, S4, S5 have controlled variations in top AlN layer thicknesses for fixed bottom 300 nm Ag / 20 nm AlN layers, while samples S2, S6, S7 have controlled variations in middle Ag layer thicknesses between fixed 20 nm AlN layers.

| Label | Top AlN layer (nm) | Ag layer (nm) | Bottom AlN layer (nm) |
|-------|--------------------|--------------| ---------------------|
| S1 | 3 | 300 | 20 |
| S2 | 20 | 300 | 20 |
| S3 | 50 | 300 | 20 |
| S4 | 100 | 300 | 20 |
| S5 | 200 | 300 | 20 |
| S6 | 20 | 25 | 20 |
| S7 | 20 | 18 | 20 |

When Ag nanoparticle diffusion and its associated LSPR phenomenon are not accounted for, standard thin film interference models (Fig. 1a) predict much weaker, broader, and more red-shifted absorption peaks (Fig. 1f,g) than those observed (Fig. 1j,k). However, the predicted interference maxima and minima (that alternate with increasing dielectric thickness) correctly dominate the ranking of absorption intensities among samples S1-S5 at around 500 nm (Fig. 1j). For detailed discussion, see Supplementary Information – Absorption Spectra.

Given that interference effects inadequately account for the observed absorption spectra, surface roughness effects need to be included in the model (Fig. 1b). Following standard practice, we introduce a roughness layer with an estimated thickness of 15 nm, as determined from the cross-sectional TEM images in Fig. 1l (Supplementary Information – Roughness Estimation). The effective index of the roughness layer is computed from the Bruggeman (BG) Effective Medium Theory (EMT) assuming a conventional[23,24] 50:50 split between Ag and AlN material contribution by volume. The corresponding scattering matrix calculated results in Fig. 1h,i are closer to the observed spectra in Fig. 1j,k. We see that the additional absorption in the visible due to the SPR associated with the rough interface is of the right order of magnitude, but the roughness model cannot produce the observed sharp peaks and tends to overestimate the absorption in the infrared. See Supplementary Information – Scattering matrix method for relevant details.

To better understand the underlying cause of the absorption, cross-sectional Scanning Electron Microscopy (SEM) images and Scanning Transmission Electron Microscopy (STEM) of sample S5 were taken in Back-Scattered Electron (BSE) mode (Fig. 1l). These images show the presence of Ag nanoparticles of ~5-10 nm radii deep into the AlN layer, which appear as high intensity specks owing to the atomic number contrast between the two materials. The LSPR excited by these Ag nanoparticles would significantly enhance the absorption properties of our samples. The resulting absorption is characteristic of both the SPR-dominated planar metal-dielectric interface (Fig. 1a,b) and the LSPR-dominated dielectric-embedded metal nanoparticles (Fig. 1e), as shown in Fig. 1c,d. The resulting hybrid structure is anticipated to have optical properties that depend on the concentration profile of Ag in the structure, as illustrated conceptually in Fig. 1c and Fig. 1d.

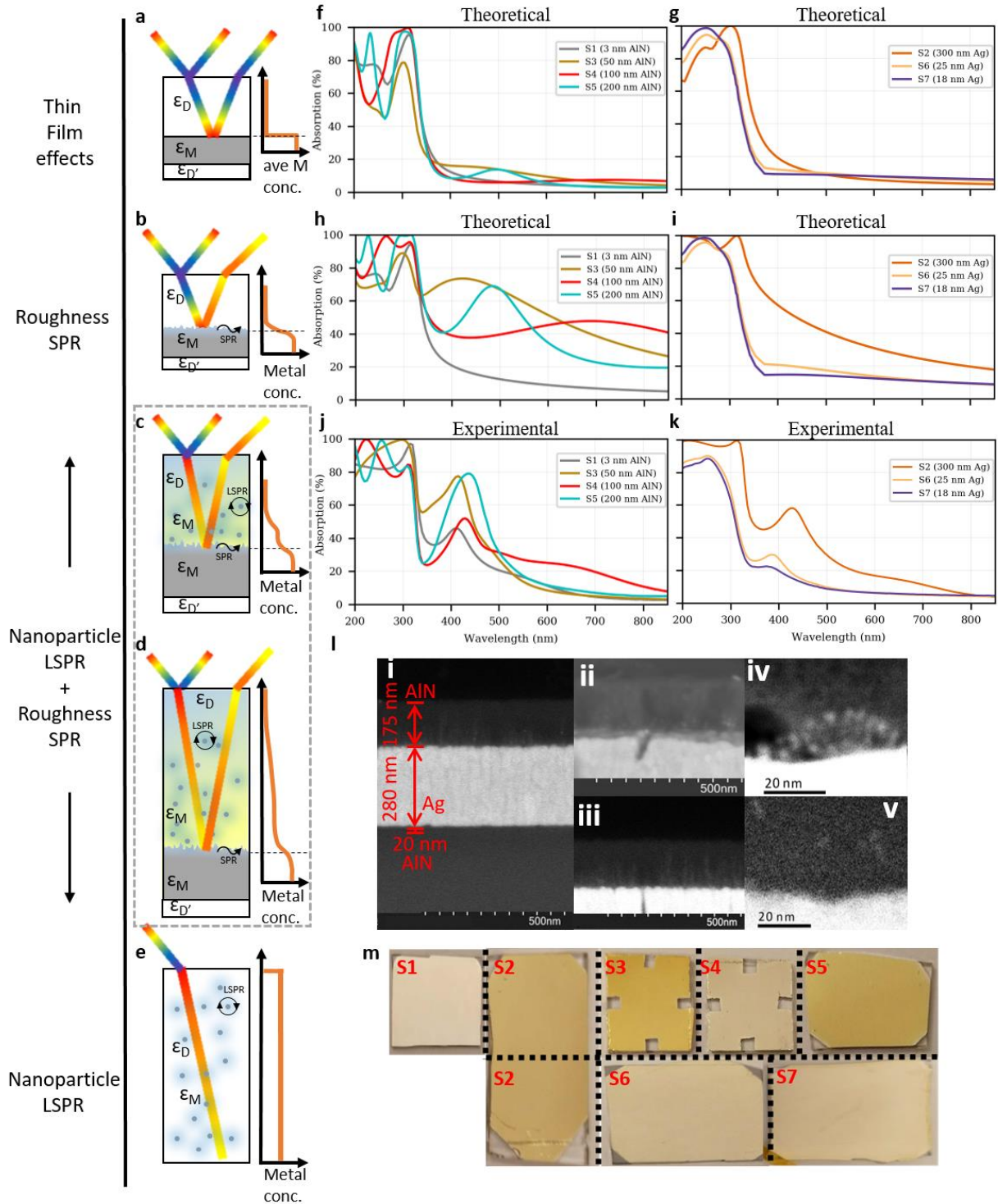

**Figure 1 | Optical role of diffused Ag nanoparticles in AlN. (a-e)** Conceptual illustrations showing the various morphological regimes which cause optical absorption in metal-dielectric films, from **(a)** the simplest thin film interference effects in the ideal metal-dielectric multilayer to **(b)** the conventional SPR from surface roughness at the metal-dielectric interface, to **(e)** a form of metal-dielectric composite with LSPR from metal nanoparticles uniformly dispersed in a dielectric, and in this study, **(c-d)** LSPR of diffused metal nanoparticles with a concentration gradient within the dielectric layer. The average metal concentration profiles are sketched on the right to summarize the central difference between the scenarios. **(f-i)** Simulated absorption spectra of the fabricated AlN / Ag / AlN multilayer samples in Table 1: assuming **(f-g)** ideal sharp interface; and **(h-i)** with additional surface roughness; **(f, h)** varying thickness of the top AlN layer on 300 nm thick Ag; and **(g, i)** varying Ag layer thickness sandwiched between 20 nm AlN layers. **(j-k)** Experimentally measured absorption spectra of samples. **(l)** (i-iii) SEM BSE cross sectional images showing high intensity specks associated with the Ag nanoparticles within the AlN; (iv-v) STEM dark field images show smaller silver nanoparticles (iv) 10-20nm away and (v) ~50nm away from the interface. **(m)** Photographs showing the colors of the samples.

## Complementary Error Function (*erfc*) ToF-SIMS depth profiles

To quantify the spatial distribution of Ag nanoparticles, Time of Flight - Secondary Ion Mass Spectroscopy (ToF-SIMS) depth profiles were taken for samples S2, S4, S5, and S6, which are shown in Fig. 2a,d,g,j, respectively. In Fig. 2, the positive secondary ion counts corresponding to the constituent elements of the film (i.e. Ag, Al, Si in AlN / Ag / AlN / Si) are plotted as a function of the ToF-SIMS sputter time, which serves as a proxy for the distance from the surface of the film. In Fig. 2a,d,g,j, we identify a region of steady ion counts as the layer of the material associated with the ion type[25]. The presence of extremely high Ag ion count, peaking where the Al ion profile undergoes drastic change, is an interfacial artefact[26–28] likely due to the steep change in oxidation states at the interface and as such does not represent a peak in the concentration of silver. Similarly, the rise in Ag ion counts near the surface of the film is likely due to local changes in the oxidation linked to oxygen contamination of the open surface. For more in-depth discussions on data interpretation, see Supplementary Information - ToF-SIMS.

We model and fit the gradients of ion counts to the complementary error function, $erfc(z)$, which is both a well-known solution to Fick's Second Law of Diffusion[29] and a commonly observed concentration profile at interfaces[28,30]. The resulting fits, shown in Fig. 2b,e,h,k and Fig. 2c,f,i,l for Ag and Al ions, respectively, match the *erfc* model closely (see Supplementary Information - ToF-SIMS for statistics). Because the respective concentrations of Ag and AlN sum up to unity, we can convert a decreasing AlN *erfc* transition to an increasing Ag *erfc* transition. By combining this converted Ag *erfc* transition with the plotted Ag *erfc* transition, we find a two-step *erfc* profile as conceptually shown in Fig. 1c,d. Given the observed Ag nanoparticle diffusion (Fig. 1l), and the natural description of diffusion concentration profiles as *erfc* functions, we attribute one of the two *erfc* functions as the concentration profile of diffused Ag nanoparticles and the other *erfc* function as the interfacial transition (including roughness) between the AlN and Ag. Then, the percentage volume of Ag, $f(x)$, at any depth $x$ is given by a linear combination of the two *erfc* functions:

$$f(x) = A_1 erfc[S_1(x - C_1)] + A_2 erfc[S_2(x - C_2)]. \qquad (1)$$

Here $A_1$ and $A_2$ are any pair of positive amplitudes that yield a physically realistic $f(x)$ that approaches 0% and 100% at the two $x$ limits. The *S* parameters control the steepness of the $erfc$ function and the *C* parameters control the centres of the $erfc$ function; together, *S* and *C* provide the linear transform $S \times (x - C)$ for the sputter time $x$, thereby enabling $erfc(x)$'s of different widths and centres. This depth profile model will be used in the upcoming optical modelling section.

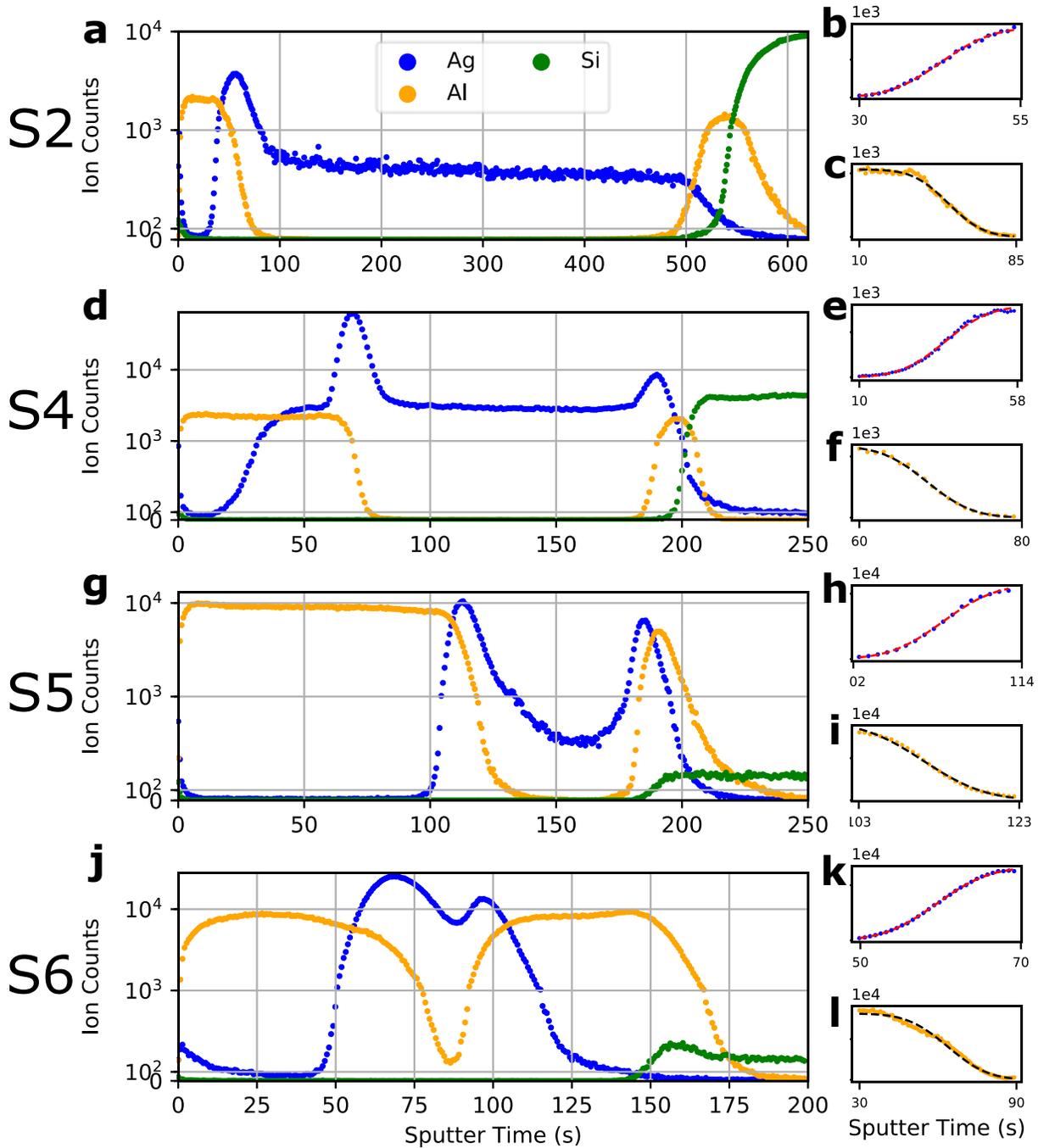

**Figure 2 | ToF-SIMS profile with analysis for samples S2, S4, S5, S6. a,d,g,j** The depth profiles of the relevant ion species. To minimize the visual impact of Ag peaks due to interface artefacts, the ion counts above 1000 are in log scale. **b-c, e-f, h-i, k-l** The regression fits to the **(b,e,h,k)** Ag ion counts and **(c,f,i,l)** to the Al ion counts, with an *erfc* model.

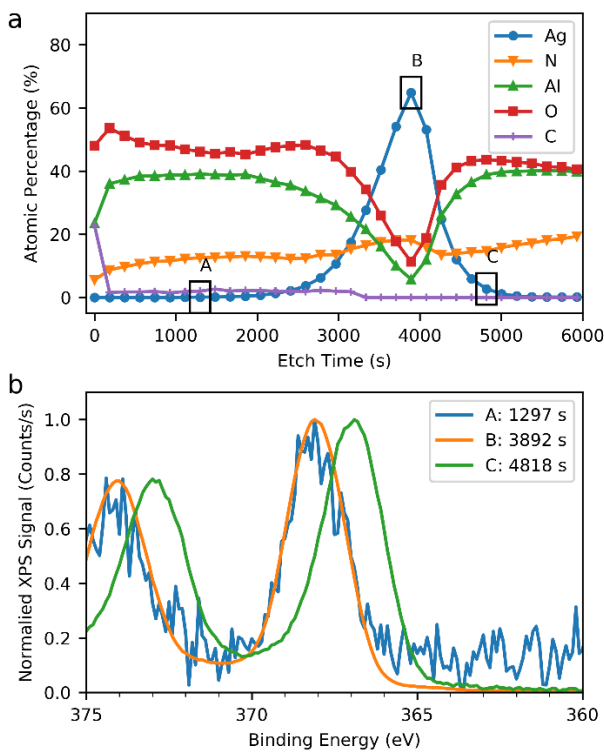

**Figure 3 | XPS Data for Sample S6. (a)** XPS depth profile of atomic concentration **(b)** XPS binding energies for the silver 3d3/2 and 3d5/2 peaks at selected depths. 1297s is within the top AlN layer, 3892s is within the silver film, and 4818s is within the seed AlN layer, close to the silver interface. The analytical probe includes contribution of signals from the top 8-10 nm of the bottom of the sputter crater. Convoluted with this are sputtering effects.

## XPS and Ag nanoparticle formation mechanism

To gain further insight on the nanoparticle formation mechanism, an XPS depth profile of sample S6 was taken. Based on the relative atomic concentrations, we calculated the Ag volume fraction as ~0.24% at an etch time of 1297s (see Supplementary Information – Calculation of Silver Volume Concentration for details). In addition, the binding energies of the silver $3d^{5/2}$ and $3d^{3/2}$ electrons provide some insight into the chemical state of silver in this structure. As can be seen in Fig. 3b, close to the interface of the bottom AlN and silver layers, the $3d^{5/2}$ and $3d^{3/2}$ peaks are at 367.0eV and 373.0eV, respectively. This is consistent with AgO[31,32]. Ag$_2$O peaks would have slightly higher energies[31,32]. Throughout the silver film and into the AlN top layer, the $3d^{5/2}$ and $3d^{3/2}$ peaks are closer to 368.1eV and 374.1eV. This is more consistent with bulk silver binding energies and therefore silver-silver bonding[31,32]. These binding energies are therefore further evidence of the silver diffusion being in the form of nanoparticles or clusters.

With the existence of Ag nanoparticles further confirmed, we can contemplate its formation mechanism. We postulate that Ag atoms thermally diffuse from the bulk Ag film surface during deposition of the top nitride layer. This is supported by the *erfc* function depth profile as well as the common knowledge that Ag is a metal with high diffusion coefficient[33] and hence diffusion can occur during the relatively low temperature sputter deposition process. With regard to the mechanism of Ag nanoparticle formation, we hypothesize that our choice of nitride dielectrics, AlN and $SiN_x$, suppresses the oxidation of silver which allows nanoparticle nucleation without the additional reduction that is usually required[34]. Here, the nitride is superior to common oxide dielectrics because the formation of $Ag_3N$ is extremely unlikely ($\Delta G_f$ =314.4 kJ/mol)[35]. Also, the high oxygen affinity of Al and Si result in preferential formation of aluminum oxide and silicon oxide which protect the diffusing silver from oxidation. Indeed, this is what we observe in the XPS measurements; we see that silver oxidation is insignificant in comparison to the significant aluminum oxide found in the multilayer.

**Calculation of optical absorption**

We now construct an optical model to account for both the diffused Ag nanoparticles and the associated LSPR as well as the rough interface and the associated SPR, for samples S3, S4, S5, and S6 (see also Samples S1 and S2 in Supplementary Information). Based on the analysis of ToF-SIMS profiles, we discretize the Ag concentration profile, $f(x)$ — as formulated in Eqn. (1), into 300 layers with increments of 0.333 vol.% Ag per discrete step. The vol.% of Ag is assumed to be constant over each discrete thickness step. We can then compute the absorption spectra of the entire multilayer using the EMT models of Maxwell-Garnett[36,37] (MG) and Bruggeman (BG). With the MG model, the effective dielectric constant $< \varepsilon >$ can be approximated as:

$$< \varepsilon > = \varepsilon_D + \frac{f \varepsilon_D (\varepsilon_M - \varepsilon_D)}{\varepsilon_D + (1-f)(\varepsilon_M - \varepsilon_D) L} \varepsilon_D$$

The depolarization coefficient $L$ accounts for the non-ideal eccentricity of the spherical inclusions. The MG model is used extensively in the modelling of metal-dielectric composites because of its success in predicting LSPR absorption[38]. In contrast, the BG model applies for large and arbitrarily shaped inclusions of fractional content $f$ such that $f$ is large compared to the values for the MG model. The resulting effective dielectric constant $< \varepsilon >$ is the solution of[39]:

$$0 = (1-f) \frac{\varepsilon_D - < \varepsilon >}{\varepsilon_D + 2 < \varepsilon >} + (f) \frac{\varepsilon_M - < \varepsilon >}{\varepsilon_M + 2 < \varepsilon >}$$

The BG model does not predict LSPR effects which are localized near the surfaces of spherical particles[38,40].

Thus, we appropriately apply the BG model to the *erfc* function representing regimes dominated by roughness induced SPR and the MG model to the *erfc* function representing regimes dominated by diffused nanoparticle induced LSPR. Fig. 4 shows the results of the regression fit in which $f(x)$ is varied to match the calculated absorption spectra with the experimental absorption spectra. For more details on the regression algorithm, see Supplementary Information – Absorption Fit from Diffusion.

The best-fit Ag concentration profiles (Fig. 4a-d) yield excellent matches with the experimental data shown in Fig. 4m-p. In Fig. 4a-c, we see that the concentration profile of diffused Ag nanoparticles, represented by the smaller *erfc* function, spreads out further from the Ag surface and assumes lower percentages in thicker top dielectric layer samples. Because thicker dielectric layers require longer sputter deposition time, there is more time for Ag close to the interface to diffuse deeper into the dielectric, as governed by Fick's Second Law[29]. Further inspection of Fig. 4a-c reveals that the effective interface between Ag and AlN, defined as the centre of the first *erfc* function from the bulk silver layer, shifts further into the dielectric layer. This is a well known phenomenon associated with the interdiffusion of two species with very different diffusion coefficients[29]. As a result, the effective thickness of the AlN layer shrinks, which causes the UV interference peaks to shift from the predicted profiles (i.e., for ideally sharp interfaces), evident from Fig. 1f and Fig. 1j.

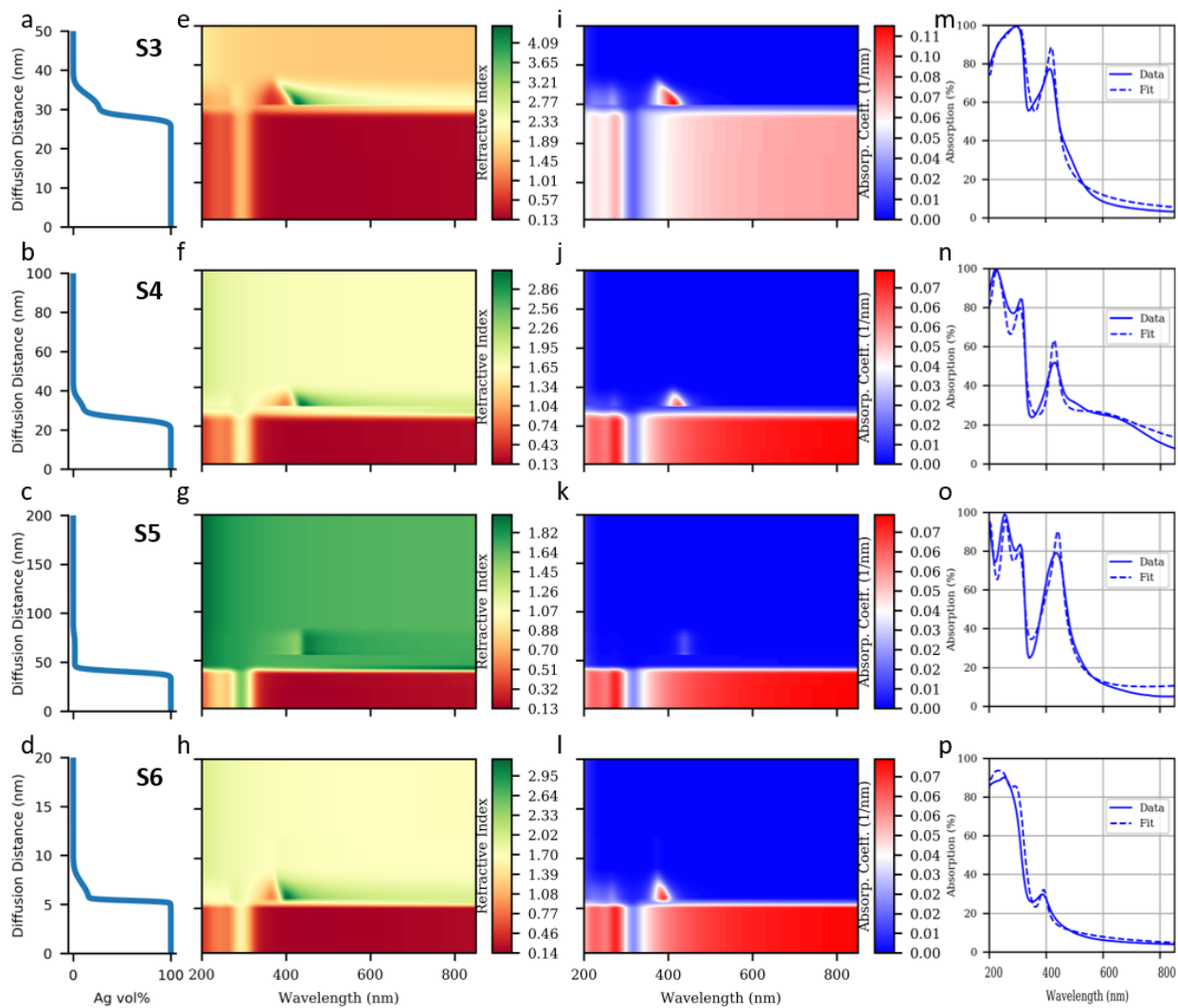

**Figure 4 | Concentration profile regression fit results for samples S3, S4, S5, S6 (rows 1 to 4, respectively). a-d** Best fit concentration profiles of Ag in the top AlN layer. The diffusion distance is the upward distance from the expected Ag layer surface. **e-h** Refractive index spectra and (**i-l**) absorption coefficient spectra obtained by BG and MG EMT in the top AlN layer, corresponding to the concentration profile. LSPR signatures are found in regions with both sharp variation in the refractive index spectrum and high absorption coefficients. **m-p** UV-Vis absorption spectra: experimental and that calculated by scattering matrix using refractive indices from EMT. Note that the diffusion distance axes have different scales and that the refractive index and absorption coefficient spectra have different color scales across the samples due to the large variations among them.

Fig. 4e-h and Fig. 4i-l show the calculated spectral refractive index and absorption coefficient profiles. The bottom of the contour plots (with silver concentration approaching 100%) corresponds to the bulk silver region, which is strongly reflective. Next, in the roughness region where the silver concentration drops significantly, it is expected that no LSPR peaks are produced because of the choice of applying the BG EMT model. Instead, the LSPR peaks are produced by the Ag in the diffusion region and are modelled by the MG EMT (see Supplementary Information – Roughness LSPR for more discussion). The LSPR peaks manifest in the violet wavelengths as sharp spectral variations in the refractive index spectrum and sharp peaks in the absorption coefficient spectrum. These LSPR peaks have intensity and width that depend strongly on the shapes of their corresponding *erfc* function concentration profile. For lower metal concentration, the MG EMT model predicts a more blue shifted and lower intensity LSPR peak[41]. Hence, as the diffusion extent increases (distance on the vertical axes of Fig. 4a-d) and the silver concentration decreases to zero, the associated band of high absorption coefficients in Fig. 4i-l blue shifts and narrows, forming an "absorption triangle" region in the distance and wavelength plane. Contrasting the two right-most columns of Fig. 4, we see that the "absorption triangle" (Fig. 4i-l) matches the shape of the experimental absorption peak (Fig. 4m-p).

For sample S5, Fig. 4c predicts that the silver diffusion profile is very extended (~30 nm into AlN) but shallow (of the order of 1%). Because of the small (~1% to 0%) variation in the diffused silver concentration in the top layer, there is less blue shifting and variation in the width of the "absorption triangle" in Fig. 4k, which translates to a relatively symmetric experimental absorption peak in Fig. 4l. Despite the lowered absorption coefficients in Fig. 4k due to low silver concentrations, the vast distance of silver diffusion dominates and leads to a very high experimental absorption peak intensity in Fig. 4o. For sample S6, the only sample in Fig. 4 with a thin Ag layer of 25 nm, we see that similar features to the other thick silver samples are exhibited, including the presence of an "absorption triangle". However, we note that the model should be further specialized to include additional parameters that account for the varied silver layer thickness due to diffusion and the interdiffusion at the second Ag / AlN interface. For more discussions related to Fig. 4, see Supplementary Information - Absorption Fit from Diffusion.

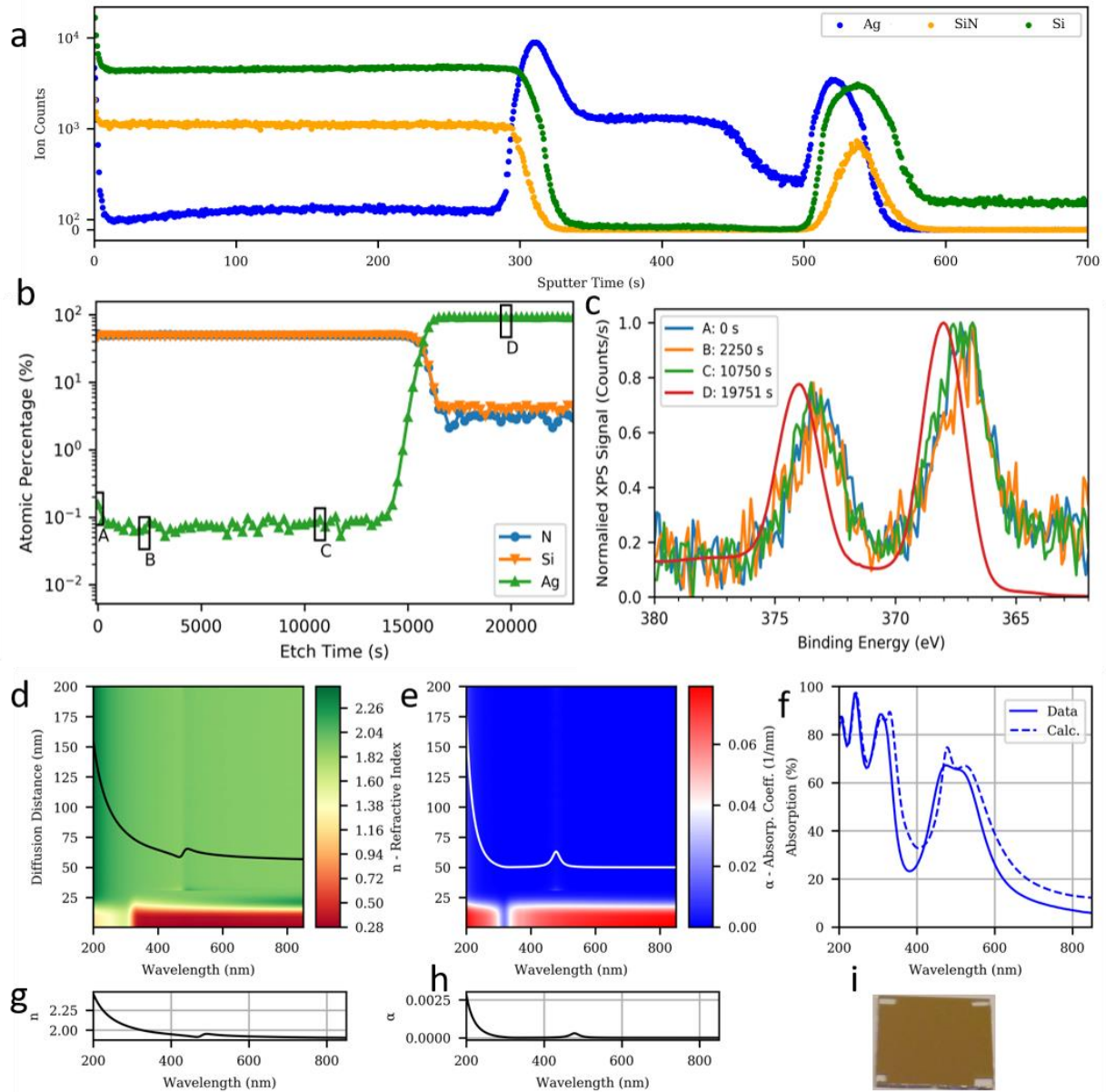

**Figure 5 | Ag nanoparticle formation in SiN$_x$. a** ToF-SIMS profile showing diffusion of Ag in the top SiN$_x$ layer. Ion counts above 1000 are in log scale. **b** XPS depth profiles of Si, N and Ag. **c** XPS Binding Energy spectra of Ag at selected points within the stack at the surface (A), subsurface (B), near interface (C), and in the Ag layer (D). There is a concentration of Ag within the SiN$_x$ layer. **d** Refractive index spectra, **e)** absorption coefficient spectra, and **f)** computed absorption spectra as obtained by MG and BG EMT in the top SiN$_x$ layer, corresponding to the concentration profile obtained from XPS. **g** Refractive index and h absorption coefficient spectrum computed from MG EMT with the representative 0.1 vol% diffused Ag nanoparticle in top SiN$_x$. LSPR occurs at ~480 nm with the anomalous index dispersion and absorption peak. Scaled versions of **g,h** are also highlighted in **d,e** as black and white lines, respectively. **i** Image of the SiN$_x$ / Ag/ SiN$_x$ stack on glass substrate.

## Ag nanoparticle formation in SiN$_x$ / Ag

We verify that the Ag nanoparticle formation in AlN / Ag / AlN is generalizable to other dielectrics by studying the SiN$_x$ / Ag / SiN$_x$ multilayer. Figure 5 shows that with ToF-SIMs (Fig. 5a) and XPS (Fig. 5b) depth profile measurements, ~0.05% a.t. (~0.1% vol.) of Ag is present within the top SiN$_x$ layer of a 200 nm SiN$_x$ / 300 nm Ag / 20 nm SiN$_x$ sputter-deposited layers (Fig. 5i). The Ag concentration profile does not follow a conventional *erfc* function profile but has a steady concentration for ~135 nm of the top SiN$_x$ thickness, which resembles the concentration profile of a limited source diffusion after a long period of time evolution[29]. As shown in Fig. 5c, the Ag 3d$^{5/2}$ and 3d$^{3/2}$ peaks in the silver film are close to 374.1 and 368.0 eV, which is similar to those observed in the Ag layer with AlN. At the surface and throughout the top nitride layer, the silver signal is noisier than in the AlN samples. However, it appears that there are additional lower energy Ag peaks (373.2 eV and 367.1 eV), along with the "bulk" silver peaks. These indicate significant oxidation of silver[31,32], and are possibly due to the outer shell of the Ag nanoparticle oxidizing[42,43].

Using the volumetric profile determined by the XPS depth profile measurement and $SiN_x$ refractive index obtained from SE (see Supplementary information- SE of $SiN_x$), we can cross validate the regression method in Fig. 4 by applying the same MG and BG EMT to directly calculate the refractive index and absorption coefficient variation over the thickness of the top $SiN_x$ as shown in Fig. 5 d,e. An absorption peak and refractive index dispersion due to the diffused silver nanoparticle LSPR is seen at 475 nm wavelength. By applying MG only to the ~0.1% vol. diffused Ag profile, and BG to the $SiN_x$ / Ag interface, we computed a double peak absorption from 375 nm to 600 nm which closely matches the experimentally determined absorption profile of Fig. 5f. The higher wavelength component of the double peak, centered at 525 nm, is due to the roughness of the Ag / $SiN_x$ interface (and its associated SPR) while the lower wavelength component, centered at 480 nm is due to the lower Ag concentration within the $SiN_x$ layer (and its associated LSPR). The broad absorption peak gives rise to a brown color in the $SiN_x$ / Ag / $SiN_x$ stack as shown in Fig. 5i.

**Conclusion**

In conclusion, we report in-situ silver nanoparticle formation at sputter-deposited Ag / dielectric (hydrogenated AlN and $SiN_x$) interfaces. The silver nanoparticles captured in SEM, STEM images have an *erfc* function spatial distribution derived from ToF-SIMS, XPS depth profiles. This creates a unique nanostructure wherein both SPR and LSPR coexist; this is especially apparent in the case of $SiN_x$ / Ag where the distinct peak wavelengths of the two plasmon modes are clearly visible in the spectra. Furthermore, we develop a physically realistic MG/BG EMT model that successfully accounts for the entire optical spectra of this nanostructure. Our findings lead to an in-situ silver nanoparticle synthesis method that seamlessly integrates with conventional fabrication of planar metal-dielectric interfaces, and thus have enormous potential for facile fabrication of devices with embedded nanoparticles for applications including light trapping[44] and SPP-LSP coupling[20–22].

## Methods

**Sputter deposition.** Hydrogenated AlN and Ag thin films of varying thickness (as summarized in Table 1), on substrates of alkaline earth boro-aluminosilicate (Corning™ Eagle XG™) 1.0-mm thick glass and (100) silicon wafer, were sputter deposited in an RF magnetron sputtering system using 99.99% pure Al and Ag targets, with 99.99% $N_2$ and Argon, and 99.9% pure $H_2$ (see Supplementary Information for details on substrate cleaning, deposition, and target pre-sputtering). The sputtering chamber was cryogenically pumped to a base pressure of less than 1.8 E-6 Torr prior to any sputtering. Ag layers were deposited using only the Ag target with Ar and $N_2$ gas flow ratio 8:8.6 sccm; AlN thin layers were deposited with reactive sputtering using the Al target with Ar, $N_2$ and $H_2$ gas flow ratio of 15:3.6:0.7 sccm. Supplementary Information Table S2 summarizes the deposition parameters of Ag and AlN thin films. The nominal thickness values of Ag films were obtained from a calibrated quartz crystal microbalance sensor. The nominal thickness and rate of deposition of AlN were determined using spectroscopic ellipsometry (SE) on SiNxgle layer AlN films deposited on (100) silicon. Further, the film thicknesses were confirmed through cross-sectional SEM. The average deposition rates were then used to fabricate the various dielectric/metal stacks. For $SiN_x$ / Ag / $SiN_x$ stacks, we reactively sputtered Si at 200W with $N_2$ at a flow ratio of 50 : 50 sccm Argon : $N_2$ for a quartz crystal sensor determined 20nm as the seed layer, sputtered Ag with $N_2$ at a flow ratio of 8 : 9 sccm Argon : $N_2$ to 300nm thickness, and reactively sputtered Si with $N_2$ at 200W to a thickness of 200nm.

**Optical measurements and images.** Spectroscopic ellipsometry (SE) measurements were taken at an angle of 60 degrees for wavelengths between 200 nm to 850nm using a SOPRA GES-5E. SE regression fit was performed with the commercial software WinElli II and the results were used to estimate the nominal film thicknesses (see Supplementary Information – SE Results for details). A Perkin Elmer Lambda 1050 UV-Vis-NIR Spectrophotometer was used to carry out the transmittance and reflectance measurements at intervals of 5 nm (see Supplementary Information – UV-Vis for details). SEM images were obtained after etching the sample cross-section with table top precision ion milling (Hitachi Ion milling system IM4000). The STEM dark field images were obtained with Hitachi HF-3300 cold field-emission. Samples were prepared via Focused Ion Beam (FIB) by first depositing a tungsten protective layer. A 40keV Ga ion beam with sequentially finer apertures was used to mill and thin the sample. After milling, but before thinning, a micro-sampling system was used to remove 3 samples of ~ 15 µm / 5 µm / 3 µm (length / height / width).

**ToF-SIMS and XPS depth profile.** ToF-SIMS depth profiles, for positive secondary ions, were obtained on ION-TOF GmbH. (Muesnter, Germany) with low energy 500 eV $O_2$ sputter beam and 30,000 eV $Bi^+$ analysis beam for sample S2, S5 and $SiN_x$ / Ag / $SiN_x$ ; and 1 keV $O_2$ sputter beam and 60,000 eV $Bi^{3++}$ analysis beam for sample S4. The crater size was 200 by 200 micron and secondary ion sampling area was 50 by 50 micron. EscaLab 250Xi XPS (Thermo Scientific, East Grinstead UK) with monochromatic AlKα X-rays (1486.68 eV) was used for the XPS analyses. The depth profile was obtained by etching the sample with $Ar^+$ ion gun at 500 eV. The Avantage software program is used to post-process the data for quantification and spectroscopic interpretation.

**Scattering matrix and EMT modelling.** Custom Python code was used to perform scattering matrix thin film multilayer calculations with BG and MG EMTs. Regression fit was performed with the Python scipy.optimize.curve_fit package.

**Data and code availability.** All the data presented herein and custom Python code used in this study are available from the corresponding author upon reasonable request.

## ACKNOWLEDGMENTS

The authors gratefully acknowledge funding from the Natural Sciences and Engineering Research Council of Canada (NSERC), the Canadian Foundation for Innovation (CFI), and the Ontario Research Fund. The authors thank Ali Zeineddine, Moein Shayegannia, Remy Ko, and Rajiv Prinja for fruitful discussions. The authors greatly appreciate characterization support by Sal Boccia as well as preliminary fabrication work by Rachel Wong Min and preliminary electrical characterization by Cindy Zhao.

## Author Contributions

Y.Y. and A.F. conceived the idea and designed the experiments. C.F., A.F. and J.L. fabricated the samples, P.B. performed ToF-SIMS, XPS, and TEM measurements, J.C., Y.Y., and A.F. performed SE measurements. J.L. and P.B. SEM imaged the samples, R.Z., Y.Y., A.F. and J.L. performed UV-Vis measurements. Y.Y. and P.B. analyzed the ToF-SIMS data, A.F., J.L., and P.B. analyzed the XPS data, J.L., and P.B. analyzed the TEM data, J.C., Y.Y., and A.F. analyzed the SE data. A.F. wrote the Python code for optical modelling, Y.Y. performed regression with the code. Y.Y., J.L., and A.F. co-wrote the principal draft of the paper with contributions from C.F., J.C., R.Z., followed by close review and editing by N.P.K., and reference management by J.C.. N.P.K. is the principal investigator.